\documentclass{amsart}
\usepackage{amsmath,amssymb,amsthm}

\newcommand{\be}{\begin{equation}}
\newcommand{\ee}{\end{equation}}
\newcommand{\bea}{\begin{array}}
\newcommand{\ea}{\end{array}}
\newcommand{\beqa}{\begin{eqnarray}}
\newcommand{\eeqa}{\end{eqnarray}}
\newcommand{\bean}{\begin{eqnarray*}}
\newcommand{\eean}{\end{eqnarray*}}

\def\up#1{\leavevmode \raise.16ex\hbox{#1}}

\newcommand{\gapproxeq}{\lower
 .7ex\hbox{$\;\stackrel{\textstyle >}{\sim}\;$}}
\newcommand{\lapproxeq}{\lower .7ex\hbox{$\;\stackrel
{\textstyle <}{\sim}\;$}}
\newcounter{appendice}

\def\thebibliography#1{{\bf REFERENCES\markboth
 {REFERENCES}{REFERENCES}}\list
 {[\arabic{enumi}]}{\settowidth\labelwidth{[#1]}\leftmargin\labelwidth
 \advance\leftmargin\labelsep
 \usecounter{enumi}}
 \def\newblock{\hskip .11em plus .33em minus -.07em}
 \sloppy
 \sfcode`\.=1000\relax}

\begin{document}
\begin{flushright}
ICN 2007-002\\
\end{flushright}
%\vspace*{5mm}
\centerline{ \LARGE   Noncommutative Field Theory from Quantum }
\centerline{\LARGE Mechanical Space-Space Noncommutativity}
\vskip .5cm
\centerline{ {\sc    Marcos Rosenbaum$^{a}$, J. David Vergara$^{b}$
and L. Roman Juarez}$^{c}$ }
\vskip 1cm
\begin{center}
Instituto de Ciencias Nucleares,\\
Universidad Nacional Aut\'onoma de M\'exico,\\
 A. Postal 70-543 , M\'exico D.F., M\'exico \\%
{\it a)mrosen@nucleares.unam.mx,\\
b)vergara@nucleares.unam.mx\\
c)roman.juarez@nucleares.unam.mx}%
\end{center}
\vskip 2cm \vspace*{5mm} \normalsize \centerline{\bf ABSTRACT} We
investigate the incorporation of space noncommutativity into field
theory by extending to the spectral continuum the minisuperspace
action of the quantum mechanical harmonic oscillator propagator with
an enlarged Heisenberg algebra. In addition to the usual
$\star$-product deformation of the algebra of field functions, we
show that the parameter of noncommutativity can occur in
noncommutative field theory even in the case of free fields without
self-interacting potentials.

\vspace{2cm}
\newpage
\section{Introduction}
Particle Quantum Mechanics can be viewed in the free field or weak
coupling limit as a minisuperspace sector of quantum field theory
where most of the degrees of freedom have been frozen. It is thus a
very convenient arena for further investigating the implications of
the quantum mechanical spacetime noncommutativity in the formulation
of field theories, as well as for evaluating the justification of
some statements that are considered as generally accepted wisdom
among practitioners of noncommutative field theory. {\it Cf e.g.}
\cite{nair} -\cite{li} and references therein for related works,
albeit in a somewhat different spirit from the problem considered
here, on noncommutative quantum mechanics. Further, in \cite{rosenb}
noncommutativity was considered within the context of the
Weyl-Wigner-Gr\"oenewold-Moyal (WWGM) formalism for extended
Heisenberg algebras, and its relation to the Bopp shift map (or what some authors
refer to as the quantum mechanical equivalent of the Seiberg-Witten
map) for expressing the algebra of extended Heisenberg operators in
terms of their commutative counterparts was discussed and results
were compared for problems previously studied in some of the above
cited works. Moreover, the canonical, noncanonical and the possible
quantum mechanical nonunitarity nature of some of these maps was
additionally analyzed in \cite{rosenba, ros1}. The results found
there are conceptually relevant to our approach here, since as shown
in several of the examples considered, transforming a problem in
NCQM into a commutative one does not always lead to two unitarily
equivalent quantum mechanical formulations.

 A case in point, arises when we compare some of our results and
their physical implications with those obtained by the procedure
followed in Ref \cite{carmona}. The comparison is quite pertinent
since both approaches are analogous in that they both have a
quantum mechanical minisuperspace as a starting point for a
construction of a field theoretical model. Indeed, the original
quantum Hamiltonian  in
\cite{carmona}, modulo some irrelevant normalizations,
 is the same as the one considered here. On the other hand, the extended original
Heisenberg algebra used in that work (Eq(2.6)) is different from
ours because the authors there require to introduce (for their
latter arguments) also noncommutativity of the momenta operators.
Making then a linear transformation (actually a Bopp shift map) to a
new set of quantum variables which satisfies the usual Heisenberg
algebra results in their new Hamiltonian (2.9). The remainder of the
construction in \cite{carmona} follows from the above. Note,
however, that the two decoupled quantum oscillators obtained in that
work are not the same as ours (to see this it suffices to set $\hat
B =0$, in their equations (2.10) and compare them with our equation
(3.60)). Thus the quantum mechanical problems implied by Eqs. (2.5)
and (2.9) in \cite{carmona} are not unitarily equivalent. In fact,
the quantum mechanical problem that is actually considered there is
that of a two dimensional anharmonic oscillator with a particular choice of
frequencies containing some constant terms labeled with the symbols
$\hat\theta$ and $\hat B$, which can not truly be identified with
the noncommutativity of any of the observables generated by the
Heisenberg algebra (2.8) characterizing the quantum problem
that at the end of the day is involved in that work. \\
Moreover, the field constructed in \cite{carmona} is a complex
scalar field, which is not so in our case, and the Feynman
propagator derived there and given in Eq(3.5) is quite different
from ours ({\it cf} Eq(\ref{equivfey})). The most important
difference being that (3.5) in Ref \cite{carmona} satisfies a highly
non-local differential equation which violates both ordinary as well
as twisted Poincar\'e invariance, while the symmetries of the Feynman
propagator we derive in this work are in agreement with recent
results on twisted NCQFT.

Based on the above  remarks and recalling that observables in
quantum mechanics are represented by Hermitian operators acting on a
Hilbert space, noncommutativity of the dynamical variables of
a quantum mechanical system can be readily understood as the noncommutativity of their
corresponding operators. In this way the physical argument that
measurements below distances of the order of the Planck length loose
operational significance \cite{dopl},  can be mathematically  described by
extending the usual Heisenberg algebra of ordinary quantum mechanics
to one including the noncommutativity of the operators related to
the spacetime  dynamical  variables. Consistently in this paper we shall
therefore use a quantum basis which is fully compatible with the noncommutativity
of the coordinates.

In particular in order to formulate space noncommutativity in Quantum Mechanics
we use the extended Heisenberg algebra with generators satisfying the commutation
relations
 \begin{eqnarray}\label{heis}
\left[\hat Q_i,\hat Q_j\right] &=& i\theta_{ij},\nonumber\\
\left[\hat Q_i,\hat P_j\right] &=& i\hbar \delta_{ij},\\
\left[\hat P_i, \hat P_j\right] &=& 0\nonumber,
\end{eqnarray}
(these could of course be generalized even more by also postulating noncommutativity of the
momenta). The parameters $\theta_{ij}$ of noncommutativity in (\ref{heis}) have
dimensions of $(length)^2$ and can, in general, be themselves
arbitrary antisymmetric functions of the spacetime operators.
However, most of the work so far appearing in the literature assumes
for simplicity that these parameters are constant, and so shall we
in what follows. The observables formed from the generators in
(\ref{heis}) act on a Hilbert space which is assumed to be the same
as the one for ordinary quantum mechanics, for any of the admissible
realizations of the extended noncommutative Heisenberg algebra.

Furthermore, utilizing the WWGM formalism for a quantum mechanical
system, with observables obeying
the above extended Heisenberg algebra,
we showed in a previous paper \cite{ros1} that the Weyl equivalent
to a Heisenberg operator $\Omega(\hat{\bf P}, \hat{\bf  Q},t)$
satisfies the differential equation
\begin{equation}\label{weyleq}
\frac{\partial\Omega_W({\bf p},{\bf q},t)}{\partial t}=-\frac{2}{\hbar}H_W \sin\left[\frac{1}{2}
(\hbar\Lambda +\sum_{i\neq j}\theta_{ij} \Lambda_{ij}')\right]\Omega_W({\bf p},{\bf q},t),
\end{equation}
where
\begin{eqnarray}\label{moyales}
\Lambda&=&\overleftarrow\nabla_{\bf q}\cdot\overrightarrow\nabla_{\bf p}-
\overleftarrow\nabla_{\bf p}\cdot\overrightarrow\nabla_{\bf q},\nonumber\\
\Lambda_{ij}'&=&\overleftarrow\partial_{q_i}\overrightarrow\partial_{q_j},
\end{eqnarray}
and $H_W$ is the Weyl equivalent to the quantum Hamiltonian.\\
Making use of (\ref{weyleq}) we further showed in \cite{ros1} that in the WWGM formalism
of quantum mechanics, the labeling variables ${\bf q}, {\bf p},$
can be interpreted as canonical classical dynamical variables
provided their algebra ${\mathcal A}$ is modified with a multiplication given
by the star-product:
\begin{equation}\label{twistedp}
q_i \star_{\theta} q_j := q_i \left(e^{\frac{1}{2}\sum_{k,l}\theta_{kl}\overleftarrow\partial_{q_k}\overrightarrow\partial_{q_l}}\right) q_j.
\end{equation}
Applying these results to the simple case of a two dimensional
harmonic oscillator satisfying the algebra (\ref{heis}), which is
taken as the unfrozen mode, or the one particle sector of a
two-component vector (or composite system) field, and using spectral analysis in order to
reconstruct the corresponding quantum field, we shall show how the
parameter of the quantum mechanical noncommutativity  appears in the
theory even for the case of a free field. This novel result, which
as we shall see is a quite natural consequence of our approach, and
contrasts with the usually made assumption that the presence of
noncommutativity in field theory is manifested only through the
deformation of the multiplication in the algebra of the fields
\cite{Seiberg}, \cite{Douglas}, \cite{Szabo}.

\section{The quantum mechanics of the harmonic oscillator in noncommutative space}

As discussed in \cite{ros1}, a configuration space basis for the quantum mechanics with an extended
Heisenberg algebra generated by (\ref{heis})  is not an
admissible basis, since the position operators $\hat Q_i, \hat Q_j,\;\;i\neq j,$ do not
simultaneously form part of a complete set of commuting observables. For  $i,j=1,2$,
then the only admissible bases for such a case are either one of the 3 sets of kets $\{|q_1, p_2\rangle\}$,
$\{|q_2, p_1\rangle\}$ and $\{|p_1, p_2\rangle\}$,
where the labels of the kets are the eigenvalues of the possible sets of commuting
observables.\\
Let us now consider the first of these bases and use the WWGM formalism and the results in \cite{ros1}
in order to evaluate the transition amplitude
$\langle q_1''(t_2), p_2''(t_2)|q_1'(t_1), p_2'(t_1)\rangle$, for a quantum 2-dimensional harmonic
oscillator with Hamiltonian
\begin{equation}\label{qham}
\hat H =\frac{1}{2m}\left(\hat P_1^2 +\hat P_2^2\right) +
\frac{m\omega^2}{2}\left(\hat Q_1^2 + \hat Q_2^2\right).
\end{equation}

>From the results in Sec. 2 of the above cited paper, it can be seen
that this transition amplitude is given by
\begin{eqnarray}\label{prop8}
\langle q_1''(t_2), p_2''(t_2)|q_1'(t_1), p_2'(t_1)\rangle &=& \langle q_1''(t_1), p_2''(t_1)|
e^{-\frac{i}{\hbar}\hat H (t_2 -t_1)} | q_1'(t_1), p_2'(t_1)\rangle \nonumber\\
         &=& \text {Tr} [\rho e^{-\frac{i}{\hbar}\hat H (t_2 -t_1)} ]\\
&=& \int dp_1 dp_2 dq_1 dq_2 \rho_W \;e^{\frac{\theta}{\hbar}p_1 \partial_{q_2}} e_{\star}^{-\frac{i}{\hbar}H_W (t_2 -t_1)},\nonumber
\end{eqnarray}
where $\theta:=\theta_{12}$,
\be\label{ro}
\rho := | q_1'(t_1), p_2'(t_1)\rangle \langle q_1''(t_1), p_2''(t_1)|,
\ee
$\rho_W$ is its corresponding Weyl function:
\be\label{row}
\rho_W = (2\pi \hbar)^{-2} \int d\xi d\eta e^{-\frac{i}{\hbar}(\eta q_2 -\xi p_1)}
\langle q_1 -\frac{\xi}{2}, p_2 - \frac{\eta}{2}|\rho|q_1 +\frac{\xi}{2}, p_2 + \frac{\eta}{2}\rangle,
\ee
and $H_W :=\frac{1}{2m}\left( p_1^2 + p_2^2\right) +
\frac{m\omega^2}{2}\left(q _1^2 + q_2^2\right)$ is the Weyl function associated with
the quantum Hamiltonian (\ref{qham}).
Substituting (\ref{ro}) into (\ref{row}) gives
\be\label{row2}
\rho_W = \frac{4}{(2\pi\hbar)^2} \delta(q_1'' +q_1' -2q_1)\delta(p_2'' +p_2' -2p_2)
\exp\left[-\frac{i}{\hbar}(2p_2 -2p_2')q_2 +\frac{i}{\hbar}(2q_1 -2q_1')p_1\right],
\ee
which, when inserted in its turn into (\ref{prop8}), yields
\begin{eqnarray}\label{prop9}
\langle q_1''(t_2), p_2''(t_2)|q_1'(t_1), p_2'(t_1)\rangle &=& \int dp_1 dq_2 e^{-\frac{i}{\hbar}(p_2'' -p_2')q_2}
\exp\{\frac{i}{\hbar}[(q_1'' - q_1') +\frac{\theta}{\hbar}(p_2'' -p_2')]p_1  \}\nonumber\\
&\times& (e_{\star}^{-\frac{i}{\hbar}H_W (t_2 -t_1)})(p_1, \frac{p_2' +p_2''}{2},\frac{q_1' +q_1''}{2}, q_2 ).
\end{eqnarray}
Note now that for an infinitesimal transition with $t_1 =t$, $t_2 =
t +\delta t$ and $q_1'' - q_1' =\dot q_1'' \delta t$, $p_2'' - p_2'
=\dot p_2'' \delta t$, (\ref{prop9}) reads \be\label{prop10} \langle
q_1''(t +\delta t), p_2''(t +\delta t)|q_1'(t), p_2'(t)\rangle =
e^{\frac{i}{\hbar}[\dot q_1' p_1 -\dot p_2'q_2 +\frac{\theta}{\hbar}
\dot p_2' p_1]\delta t} e^{-\frac{i}{\hbar}H_{cl}(p_1, p_2', q_1',
q_2)\delta t} , \ee where $H_{cl}$ (= $H_W$ for the case here
considered) is the classical Hamiltonian resulting from making the
replacements $\hat Q \rightarrow q$ and $\hat P\rightarrow p$ in the
original quantum Hamiltonian (\ref{qham}). Following Feynman's path
integral formalism, the transition over a finite time interval is
then given by \be\label{prop11} \langle q_1''(t_2),
p_2''(t_2)|q_1'(t_1), p_2'(t_1)\rangle \sim \int {\mathcal D}q_1
{\mathcal D}p_2 {\mathcal D}p_1 {\mathcal D}q_2 \exp
\{\frac{i}{\hbar}\int_{t_1}^{t_2} [\dot q_1 p_1 -\dot p_2 q_2
+\frac{\theta}{\hbar} \dot p_2 p_1 - H_{cl}] dt \}. \ee This result
(for an alternate derivation see \cite{acat} and related work in
\cite{Mangano}-\cite{drag}) provides an univocal procedure for
obtaining the Feynman propagator in spacetime noncommutative quantum
mechanics as well as the expression for the deformed classical
action, which in our particular case is given by
\begin{equation}\label{piact1}
S(q_1, p_2, q_2, p_1,t)=\int_{t_1}^{t_2}[\dot q_1 p_1 -\dot p_2 q_2 +\frac{\theta}{\hbar} \dot p_2 p_1 -H_{cl} ]dt.
\end{equation}

Let us next re-write the action (\ref{piact1}) in the form
\be\label{fpiact}
S=\int dt\left[p_1 \dot{q}_1 -\dot{p}_2(q_2 -\frac{\theta}{\hbar}p_1)
 -\frac{{p_1}^2}{2m}-\frac{{p_2}^2}{2m}-\frac{m\omega^2}{2}{q_1}^2-\frac{m\omega^2}{2}{q_2}^2 \right],
\ee
which, when setting
\be\label{ccv}
\tilde{q}_2 = q_2 -\frac{\theta}{\hbar}p_1,
\ee
 results in
\be\label{redact1}
S=\int dt\left[p_1 \dot{q}_1 -\dot{p}_2\tilde{q}_2
 -\frac{{p_1}^2}{2m}-\frac{{p_2}^2}{2m}-\frac{m\omega^2}{2}{q_1}^2-\frac{m\omega^2}{2}(\tilde{q}_2 +\frac{\theta}{\hbar}p_1)^2
 \right].
\ee
Fixing now $p_1$ and $\tilde{q}_2$ at the end points and varying with respect to these variables
we get
\begin{eqnarray}
\dot{q}_1 &=& \frac{p_1}{m}+\frac{m^2\omega^2 \theta}{\hbar}(\tilde{q}_2 +\frac{\theta}{\hbar}p_1) ,\\
\dot{p}_2 &=& -m\omega^2(\tilde{q}_2 +\frac{\theta}{\hbar}p_1).
\end{eqnarray}
>From the above we derive
\begin{eqnarray}
p_1&=& m\dot{q}_1 +\frac{m \theta}{\hbar}\dot{p}_2 \\
\tilde{q}_2 &=& -\frac{1}{m\omega^2}[\dot{p}_2 +\frac{m^2\omega^2 \theta}{\hbar}(\dot{q}_1 +\frac{\theta}{\hbar}\dot{p}_2)].
\end{eqnarray}
Substituting these last expressions into (\ref{redact1}) shows  that (\ref{prop11}) may be reduced to
\begin{equation}\label{prop2}
\langle q_1''(t_2), p_2''(t_2)|q_1'(t_1), p_2'(t_1)\rangle\sim\int\int {\mathcal D}q_1{\mathcal D}p_2
e^{\frac{i}{\hbar} S(q_1,p_2,t)},
\end{equation}
with
\begin{equation}\label{action}
S(q_1,p_2,t)=\int dt\left[ \frac{m}{2}\dot q_1^2 +\frac{m\theta}{\hbar}\dot p_2 \dot q_1 +
\left( \frac{1}{2m\omega^2} +\frac{m\theta^2}{2\hbar^2}\right)\dot p_2^2 -\frac{p_2^2}{2m}
-\frac{m\omega^2}{2}q_1^2\right].
\end{equation}

Note that by varying (\ref{action}), it follows that the canonical dynamical variables $q_1$ and $p_2$ obey the
set of second order coupled ordinary differential equations
\begin{equation}\label{cdeq}
\left( \begin{array}{c} \ddot {q}_1 \\ \ddot {p}_2 \\ \end{array}\right) =-
\left( \begin{array}{*{20}c}
\frac{m^2 \omega^4 \theta^2}{\hbar^2} +\omega^2 & -\frac{\omega^2 \theta}{\hbar}\\
-\frac{m^2 \omega^4 \theta}{\hbar} & \omega^2\\  \end{array}\right)
\left( \begin{array}{c}  q_1 \\ p_2 \\ \end{array}\right),
\end{equation}
which, when diagonalized, decouple into two harmonic oscillators with frequencies given by
\begin{equation}\label{omegas}
\omega_{1,2}= \omega\left[1+\frac{m^2 \omega^2 \theta^2}{2\hbar^2}\pm
\frac{m\omega\theta}{2\hbar}\sqrt {4+\frac{m^2 \omega^2 \theta^2}{\hbar^2}}\right]^\frac{1}{2}.
\end{equation}
Hence, the energy eigenvalues of (\ref{qham}) are
\begin{equation}\label{eeigen}
E=\hbar\omega_1 (n_1+\frac{1}{2}) +\hbar\omega_2 (n_2+\frac{1}{2}).
\end{equation}

It is pertinent to emphasize here that the change of variables
at the classical level involved in Eq. (\ref{ccv})  does not correspond to
a Bopp shift, so it also does not follow that making such a change of
variables in the action (\ref{fpiact}) implies that we are passing from NCQM
to ordinary quantum mechanics. \\

\subsection{Hamiltonian formulation}
Consider now the Lagrangian $L$ in the action (\ref{action}) and make the identifications
\begin{equation}\label{zetas}
z_1:=q_1,\;\;\;\; z_2:=\frac{p_2}{m\omega},
\end{equation}
so that both $z_1$ and $z_2$ have dimension of length. Furthermore,
introducing the dimensionless quantity $\tilde\theta$ :
\begin{equation}\label{tildteta}
\tilde\theta=\frac{m\omega\theta}{\hbar},
\end{equation}
with $m,\omega$, being some characteristic mass and frequency,
respectively, to be further specified below, we can then write
\begin{equation}\label{zlag}
L=\frac{1}{2}\left[ \dot z_1^2 -\omega^2 z_1^2
+\dot z_2^2 -\omega^2 z_2^2
+2\tilde\theta \dot z_1 \dot z_2 +\tilde\theta^2 \dot z_2^2 \right].
\end{equation}
The momenta canonical to the $z_i$'s are
\begin{eqnarray}\label{canmom}
\pi_{1} &=& \dot z_1 +\tilde\theta \dot z_2, \nonumber\\
\pi_2 &=& \dot z_2 +\tilde\theta \dot z_1 +\tilde\theta^2 \dot z_2.
\end{eqnarray}
Inverting (\ref{canmom}) we have
\begin{equation}\label{inv}
\left( \begin{array}{c} \dot {z}_1 \\ \dot {z}_2 \\ \end{array}\right) =
\left( \begin{array}{*{20}c}
1 +\tilde\theta^2 & -\tilde\theta\\
-\tilde\theta & 1\\  \end{array}\right)
\left( \begin{array}{c}  \pi_1 \\ \pi_2 \\ \end{array}\right),
\end{equation}

from where it follows that
\begin{equation}\label{zham}
H=\pi_1 \dot z_1 +\pi_2 \dot z_2 -L= \frac{1}{2}\left [ (1+\tilde\theta^2)\pi_1^2 +\pi_2^2 -2\tilde\theta
\pi_1 \pi_2 +\omega^2 z_1^2 + \omega^2 z_2^2 \right].
\end{equation}

Making use of the theory of quadrics we can diagonalize (\ref{zham}) by first solving for the eigenvalues
$\lambda_{1,2}$ of the characteristic determinant of the matrix
$$\left( \begin{array}{*{20}c}
\frac{1}{2}(1 +\tilde\theta^2) & -\frac{\tilde\theta}{2}\\
-\frac{\tilde\theta}{2} & \frac{1}{2}\\  \end{array}\right). $$
We thus get
\be\label{lambdas}
\lambda_{1,2}=\frac{1}{2} \left(1+\frac{\tilde\theta^2}{2} \pm \frac{\tilde\theta}{2}\sqrt {4+\tilde\theta^2} \right).
\ee
Hence
\begin{eqnarray}\label{zham2}
H&=& \left(\pi_1, \pi_2 \right)\left(\tilde M \right) \left( M \right)\left( \begin{array}{*{20}c}
\frac{1}{2}(1 +\tilde\theta^2) & -\frac{\tilde\theta}{2}\\
-\frac{\tilde\theta}{2} & \frac{1}{2}\\  \end{array}\right)\left(\tilde M \right) \left( M \right)
\left( \begin{array}{c}  \pi_1 \\ \pi_2 \\ \end{array}\right)+\frac{\omega^2}{2}\left(  z_1^2 +  z_2^2 \right)\nonumber\\
&=&(\pi'_1, \pi'_2)\left( \begin{array}{*{20}c}
\lambda_1 & 0\\0 & \lambda_2\\  \end{array}\right)\left( \begin{array}{c}  \pi'_1 \\ \pi'_2 \\ \end{array}\right)
+\frac{\omega^2}{2}\left(  z'^2_1 +  z'^2_2 \right),
\end{eqnarray}
where \be \left( \begin{array}{c}  \pi'_1 \\ \pi'_2 \\
\end{array}\right)= \left(M\right)\left( \begin{array}{c}  \pi_1 \\
\pi_2 \\ \end{array}\right),\;\;\; \left( \begin{array}{c}  z'_1 \\
z'_2 \\ \end{array}\right)= \left(M\right)\left( \begin{array}{c}
z_1 \\ z_2 \\ \end{array}\right), \ee
and the entries of the
symmetric matrix $\left(M\right)=\left( \begin{array}{*{20}c} m_{11}
&m_{12} \\m_{21} & m_{22}\\  \end{array}\right)$ are given by
\be\label{orthom1}
m_{11}=-\frac{1}{\sqrt{1+\frac{\omega_2^2}{\omega^2}}}, \;\;\;\;\;
 m_{12}=-\frac{\frac{\omega_2^2}{\omega^2}-1}{\tilde\theta \sqrt{1+\frac{\omega_2^2}{\omega^2}}},
\ee
\be
m_{21}=\frac{1}{\sqrt{1+\frac{\omega_1^2}{\omega^2}}}, \;\;\;\;
 m_{22}=\frac{\frac{\omega_1^2}{\omega^2}-1}{\tilde\theta \sqrt{1+\frac{\omega_1^2}{\omega^2}}}
\ee
where, by using (\ref{omegas}) and (\ref{tildteta}), one can readily verify that $m_{12} =m_{21}$
as required.\\
If we finally let \be z'_i =(\lambda_i)^{\frac{1}{2}}
x_i,\;\;\;\;\; \pi'_i =(\lambda_i)^{-\frac{1}{2}} \pi_{x_i},\;\;
i=1,2, \ee we arrive at \be\label{diagham} H=\pi^{2}_{x_1}
+\pi^{2}_{x_2} +\frac{1}{4}\left({\omega_1}^2 x^{2}_1 +\omega_2^2
x^{2}_2\right). \ee

It should be clear from the above calculations that the transformed
variables $x_i,\pi_{x_i}$ remain canonically conjugate to each other.
Thus it follows from the Hamilton equations that \be\label{hame}
\pi_{x_i} =\frac{1}{2} \dot x_i, \ee so the Lagrangian (\ref{zlag})
now reads \be\label{diaglag} L=\frac{1}{4}\left(\dot x^{2}_1 +\dot
x^{2}_2 -\omega_1^2 x^{2}_1 - \omega_2^2 x^{2}_2\right). \ee
Variation of this expression with respect to $x_i$ yields
\be\label{ho} \ddot x_i +\omega_i^2 x_i=0, \ee which are indeed the
equations of motion for two decoupled harmonic oscillators with
respective frequencies $\omega_i$, as asserted previously.\\
Furthermore, it can be readily verified that the point
transformations \be\label{pt1} \left( \begin{array}{c}  \pi_{x_1} \\
\pi_{x_2} \\ \end{array}\right)= \left( \begin{array}{*{20}c}
m_{11}\sqrt{\lambda_1} &m_{12}\sqrt{\lambda_1} \\m_{21}\sqrt{\lambda_2} & m_{22}\sqrt{\lambda_2}\\
 \end{array}\right)\left( \begin{array}{c}  \pi_1 \\ \pi_2 \\ \end{array}\right),
\ee and \be\label{pt2} \left( \begin{array}{c}  x_1 \\ x_2 \\
\end{array}\right)= \left( \begin{array}{*{20}c}
\frac{m_{11}}{\sqrt{\lambda_1}} &\frac{m_{12}}{\sqrt{\lambda_2}}
\\\frac{m_{21}}{\sqrt{\lambda_1}} & \frac{m_{22}}{\sqrt{\lambda_2}}\\
 \end{array}\right)\left( \begin{array}{c}  z_1 \\ z_2 \\ \end{array}\right),
\ee are canonical, with generating function \be\label{gf}
F_2(z_1,z_2, \pi_{x_1}, \pi_{x_2}) = \sum_{i,j}
\frac{m_{ij}}{\sqrt{\lambda_i}} z_j \pi_{x_i}. \ee We also have that
when substituting (\ref{pt2}) into the Lagrangian (\ref{diaglag}) we
recover (\ref{zlag}) and that the Jacobian of each the
transformations (\ref{pt1}) and (\ref{pt2}) is equal to
$\frac{1}{2}$, so that
$${\mathcal D} x_1 {\mathcal D} x_2 =\frac{1}{2}{\mathcal D} z_1 {\mathcal D} z_2.$$
Consequently, the quantum mechanics derived from the path integral with the action
(\ref{action}) is unitarily equivalent to the path integral formulation based on the
action resulting from the diagonalized Lagrangian (\ref{diaglag}).

\section{Field theoretical model}

Paralleling standard quantum field theory we next construct a noncommutative
field theory over a $(1+2)$-Minkowski space by taking
an infinite superposition of the quantum mechanical harmonic oscillator minisuperspaces
described by (\ref{diaglag}).
%Paralleling this procedure, we shall  show next how a noncommutative
%field theory over a $1+2$ Minkowski space, with the spatial
%coordinates, and by extension the fields, obeying
%the twisted algebra (\ref{twistedp}), can be derived by considering
%a mechanical system of infinitely many degrees of freedom,
Each of these oscillators  consists of the pair $x_1({\bf k}),
x_2({\bf k})$, labeled by the continuous parameter $\bf k$ and
satisfying (\ref{ho}). Thus in our construction, the quantum
mechanical spacial noncommutativity will reflect itself both in the
deformation parameter dependence of the different frequencies of the
pairs of oscillators, as well as in the twisting of the product of
the algebra of the resulting fields. Consequently this simple model
shows that spacetime noncommutativity can be present in field theory
even in the absence of self-interaction potentials.\\

Let us consider a field system $\Phi_i({\bf q}, t), \;i=1,2$,
over a $(1+2)$-Minkowski space-time, satisfying the uncoupled Klein-Gordon field equations
\begin{equation}\label{feq}
\left( {\begin{array}{*{20}c}
   \square^2 +\mu_1^2 & 0  \\
   0 & \square^2 +\mu_2^2 \\
\end{array} } \right)\left( {\begin{array}{c} \Phi_1(q_1,q_2,t)\\ \Phi_2(q_1,q_2,t)\\ \end{array} }\right)=0 ,
\end{equation}
where
\begin{equation}\label{ft}
\Phi_i({\bf q}, t) = (2\pi)^{-1} \int d{\bf k} \;x_i({\bf
k},t)\;e_{\star_\theta}^{i\bf k.\bf q} ,
\end{equation}
and
\be\label{ft2}
e_{\star_\theta}^{i\bf k.\bf q}:= 1+ i{\bf k.\bf q} + \frac{1}{2}(i\bf {k. q})\star_{\theta}(i\bf {k. q}) +\dots.
\ee
Note that in the above definition of the field system in terms of
its Fourier transform we have used the star-exponential for
describing plane waves. Our rationale for this is based on the
observation made in (\cite{rosenb}) where, by making use of the WWGM
formalism and elements of quantum group theory, we show that quantum
noncommutativity of coordinate operators in the extended Heisenberg
algebra leads to a deformed product of the classical dynamical
variables that is inherited to the level of quantum field theory.
This deformed product is the so called Moyal star-product defined in
(\ref{twistedp}). Thus, expressing the fields as in (\ref{ft})
guarantees explicitly that they are elements of the deformed algebra
${\mathcal A}_\theta$ with the $\star$-multiplication.

Note also that in (\ref{feq}) the D'Alembertian is given by
\begin{equation}\label{square}
\square^2=\partial_t^2 -\bar\partial_i^\dag\bar\partial_i,
\end{equation}
with the anti-hermitian derivation $\bar\partial_i$ defined by \cite{gaumme}:
\begin{equation}\label{deriv}
\bar\partial_i=\theta_{ij}^{-1}\;ad_{q_j},\;\;\;\text {and}\;\; \bar\partial_i^\dag=-\bar\partial_i ,\;\;\;i=1,2,
\end{equation}
and where the adjoint action is realized by the twisted product commutator
\begin{equation}\label{twist2}
[q_i, q_j]_{\star_\theta}:=q_i \star_\theta q_j -q_j \star_\theta q_i.
\end{equation}
Thus,  the algebra (\ref{twistedp}) has been incorporated into (\ref{feq})
%, both in the frequencies $\omega_i$ as well as
through the defining Fourier transformation equation (\ref{ft}) for the fields since
these, as functions of the $q_i$'s, they
inherit the ${\star}$-multiplication and are therefore also elements of the twisted algebra $\mathcal A_\theta$.

Now, by making use of the Baker-Campbell-Hausdorff theorem, together with the commutator (\ref{twist2})
as well as of the identity $[q_2, q_1^n]_{\star_\theta}=-in\theta q_1^{(n-1)}$, we
have that
\begin{eqnarray}\label{rel1}
\bar\partial_1(e_{\star_\theta}^{i\bf k.\bf q}) &=&\theta^{-1}[q_2,e^{ik_1 q_1}e^{ik_2 q_2}
e^{\frac{i}{2}k_1 k_2\theta}]_{\star_\theta}\nonumber\\                                 &=&k_1e_{\star_\theta}^{i\bf k.\bf q},
\end{eqnarray}
and (recalling that $\bar\partial_i^\dag =-\bar\partial_i$)
\begin{equation}\label{rel2}
\bar\partial_1^\dag\bar\partial_1(e_{\star_\theta}^{i\bf k.\bf q})=-k_1^2 e_{\star_\theta}^{i\bf k.\bf q}.
\end{equation}
Similarly
\begin{equation}
\bar\partial_2^\dag\bar\partial_2(e_{\star_\theta}^{i\bf k.\bf q})=-k_2^2 e_{\star_\theta}^{i\bf k.\bf q}.
\end{equation}
We therefore find that the field equations (\ref{feq}) read
\begin{equation}\label{feq2}
(\square^2 +\mu_i^2)\Phi_i({\bf q}, t) = (2\pi)^{-1}\int d{\bf k}
[\ddot {x}_i+({\bf k}^2 +\mu_i^2)x_i ] e_{\star\theta}^{i\bf k.\bf
q}=0, \;\;i=1,2.
\end{equation}
Using next the orthonormality
\begin{equation}\label{orton}
(2\pi)^{-2}\int\int dq_1dq_2 e_{\star_\theta}^{i\bf k.\bf q}e_{\star_\theta}^{-i\bf k'.\bf q}=\delta(\bf k -\bf k'),
\end{equation}
and the dispersion relation
\begin{equation}\label{disp}
 {\bf k}^2 +\mu_i^2=k_0^2=\omega_i^{2}(\bf k) ,
\end{equation}
we obtain from the right hand of (\ref{feq2}):
\begin{equation}\label{rel3}
\ddot x_i({\bf k},t) +\omega_i^{2}(\bf k) x_i({\bf k},t) =0.
\end{equation}
Observe that $\omega_i(\bf k)$, $i=1,2$, in (\ref{disp}) is given by
(\ref{omegas}) with $\omega\rightarrow \omega(\bf k)$ and
$\theta\rightarrow \theta(\bf k)$ being now respectively the wave vector
dependent frequency in (\ref{heis}) and the noncommutative  parameter
of the quantum mechanical system for each $\bf k$ in the spectral
decomposition (\ref{ft}).
Comparing (\ref{rel3}) with (\ref{ho}), and observing that according
to our definition (\ref{tildteta}) we now have $\theta({\bf
k})=\frac{\hbar\tilde\theta}{m\omega(\bf k)}$ , we
choose $\theta(\bf k)$ such that $\tilde\theta$ remains a pure number
independent of ${\bf k}$. We then have that the Lagrangian
(\ref{diaglag}) for the pair of decoupled harmonic oscillators
$x_i({\bf k},t)$ can be seen, for a fixed value of the continuum
parameter $\bf k$, as a minisuperspace of the full field theory
characterized by the action:
\begin{equation}\label{eqfaction2}
\begin{split}
S=\int dtdq_1dq_2 \;{\mathcal L}=\frac{1}{2}\int dtdq_1dq_2\left[\dot\Phi_1^\dag \star_{\theta}\dot\Phi_1-
(\bar\partial_i\Phi_1)^\dag \star_{\theta}\bar\partial_i\Phi_1 -\mu_1^2\Phi_1^\dag \star_{\theta}\Phi_1  \right.\\
\left.+\dot\Phi_2^\dag \star_{\theta}\dot\Phi_2 - (\bar\partial_i\Phi_2)^\dag \star_{\theta}
\bar\partial_i\Phi_2 -\mu_2^2\Phi_2^\dag \star_{\theta} \Phi_2 +\frac{1}{2}(\Phi_1^\dag\star_{\theta} J_1({\bf q},t)\right.\\
\left.  + J_1^\dag({\bf q},t)\star_{\theta} \Phi_1) +
\frac{1}{2}(\Phi_2^\dag \star_{\theta}J_2({\bf q},t) + J_2^\dag({\bf q},t)\star_{\theta} \Phi_2)\right],
\end{split}
\end{equation}
after adding two arbitrary external driving sources.\\
Note that in the above expression we have formally included the $\star$-product for the
algebra of the fields, even though in fact, in the absence of field interaction potentials, these could be ignored
in view of the identity
\begin{equation}\label{ident2}
\int dq_1dq_2 f({\bf q})\star_\theta g({\bf q})=\int dq_1dq_2 f({\bf q}) g({\bf q}),
\end{equation}
which follows directly by parts integration. However, also note that the noncommutativity parameter
$\tilde \theta$ will still be present in the frequencies $\omega_i(\bf k)$ even in such a case, since these now
read
\begin{equation}\label{omegas2}
\omega_{1,2}({\bf k})= \omega({\bf k})\left[1+\frac{\tilde\theta^2}{2}\pm
\frac{\tilde\theta}{2}\sqrt {4+\tilde\theta^2}\right]^\frac{1}{2}.
\end{equation}

\section{Path integral and Feynman propagator}
In order to derive the Feynman propagator for our theory,
we use (\ref{ft}) and a similar expression for the Fourier transform $\tilde F_i$ of the sources $J_i$ together
with (\ref{orton}), as well as the transformations
\begin{eqnarray}\label{ft3}
x_i(({\bf k}, t) &=& (2\pi)^{-\frac{1}{2}} \int dk_0 e^{ik_0 t} {\tilde x}_i({\bf k}, k_0),\nonumber\\
F_i(({\bf k}, t) &=& (2\pi)^{-\frac{1}{2}} \int dk_0 e^{ik_0 t} \tilde F_i({\bf k}, k_0).
\end{eqnarray}
We thus get
\begin{equation}\label{act3}
\begin{split}
S= \frac{1}{2} \int dk_0 d{\bf k}\left[(k_0^2 -{\bf k}^2
-\mu^2)\left(\sum_{i=1,2} {\tilde x}_i({\bf k}, k_0)
{\tilde x}_i({\bf k}, -k_0)\right) \right.\\
+\left.{\tilde x}_1({\bf k}, k_0)\tilde F_1({\bf k}, -k_0)+{\tilde
x}_1({\bf k}, -k_0)
\tilde F_1({\bf k}, k_0)\right.\\
+\left.{\tilde x}_2({\bf k}, k_0)\tilde F_2({\bf k}, -k_0)+{\tilde
x}_2({\bf k}, -k_0) \tilde F_2({\bf k}, k_0)\right ].
\end{split}
\end{equation}

Following standard procedures (see {\it e.g.} \cite{ramond}), we now make the change of variables
\begin{eqnarray}\label{cv}
{\tilde x}_1({\bf k}, k_0)&=& Z_1({\bf k}, k_0)+\beta(k_0)\tilde F_1({\bf k}, k_0)+\gamma(k_0)F_2({\bf k}, k_0),\nonumber\\
{\tilde x}_2({\bf k}, k_0)&=& Z_2({\bf k}, k_0)+\lambda(k_0)\tilde
F_1({\bf k}, k_0)+\nu(k_0)F_2({\bf k}, k_0).
\end{eqnarray}
Inserting (\ref{cv}) into (\ref{act3}) and requiring that terms linear in the $Z_i$'s cancel,
allows us to fix the parameters $\beta, \gamma, \lambda, \nu$ as:
\begin{eqnarray}\label{param}
\beta(k_0)&=& (k_0^2 -{\bf k}^2 -\mu^2)^{-1},\nonumber\\
\lambda(k_0)&=&\gamma(k_0)=0,\nonumber\\
\nu(k_0)&=&-{ (k_0^2 -{\bf k}^2 -\mu^2)}^{-1}.
\end{eqnarray}
If we next replace (\ref{param}) into the action resulting from (\ref{act3}) by the above procedure, we
derive the following contribution to the integrand in that action from the terms quadratic in the sources:
\begin{equation}\label{equivfey8}
\begin{split}
\langle Z_0[J]\rangle:=
-\frac{1}{2}\int\dots\int d{\bf q }\; d{\bf q' }\; dt\;dt'
\left(  J_1^\dag({\bf q}, t) \;\;\; J_2^\dag({\bf q}, t )\right)\hspace{1in}\\
\times\left( {\begin{array}{*{20}c}
  D_1({\bf q} -\bf q', t-t')& 0 \\
   0 & D_2({\bf q} -\bf q', t-t') \\\end{array} } \right)
\left( {\begin{array}{c} J_1({\bf q'},t')\\  J_2({\bf q'}, t')\\ \end{array} }\right),
\end{split}
\end{equation}
where $D_i({\bf q}-{\bf q}',t-t')$ are the Feynman propagators:
\begin{equation}\label{equivfey}
 D_i({\bf q}-{\bf q}',t-t')=(2\pi)^{-3}\int\dots\int d{\bf k}\; dk_0
\left(\frac{e^{-i[k_0(t-t')-{\bf k}.({\bf q}-{\bf q}')]}}{k_0^2 -\omega_i^{2}({\bf k}) +i\epsilon}\right), \;\;\;i=1,2,
\end{equation}
and the $\omega_i^{2}({\bf k})$ are given by (\ref{omegas2}).

Note that these propagators satisfy the Klein-Gordon equations
\begin{equation}\label{tklein}
\left(\square^2+\mu_i^2 \right) D_i({\bf q}-{\bf q}',t-t')=-\delta({\bf q}-{\bf q}')
\delta(t-t').
\end{equation}
 Observe also that (\ref{tklein}) is invariant under the twisted Poincar\'e transformations
 discussed in \cite{chaichian2}, since
the D'Alembertian, as defined in (\ref{square}), is invariant under
these transformations and the indices $i=1,2$, are not space-time indices.\\
In consequence of the above, the vacuum to vacuum amplitude for our theory is thus given by
\begin{equation}\label{amp}
W[J]=W[0] e^{\frac{i}{\hbar}\langle Z_0[J]\rangle},
\end{equation}
and the classical fields $\Phi^{(0)}_{(cl)i} \equiv -i\frac{\delta \ln W_0}{\delta J_i^{\dag}({\bf q},t)}
=\frac{\delta Z_0}{\delta J_i^\dag({\bf q},t)}$
 satisfy the driven Klein-Gordon field equations
\begin{equation}\label{K-G}
(\square^2 +\mu_i^2)\Phi^{(0)}_{(cl)i} =\frac{1}{2}J_i.
\end{equation}

\section{Second Quantization }

Let us  promote the $x_i$ in (\ref{ft}) to the rank of operators
and, similarly to that equation, let us define field canonical
momenta by \be\label{ft21} \hat\Pi_i =(2\pi)^{-1} \int d{\bf k}
\;{\hat\pi}_i({\bf k},t)\;e_{\star_\theta}^{-i\bf k.\bf q} , \ee
with $\hat x_i, \hat\pi_j$, satisfying now the commutation relations
\begin{eqnarray}\label{cancomm}
\left[\hat x_i({\bf k},t), \hat\pi_j({\bf k'},t)\right]&=&i\hbar\delta_{ij}\delta({\bf k}-{\bf k'}),\nonumber\\
\left[\hat x_i({\bf k},t), \hat x_j({\bf k'},t)\right]&=&0,\\
\left[\hat \pi_i({\bf k},t), \hat\pi_j({\bf
k'},t)\right]&=&0.\nonumber
\end{eqnarray}
Assuming further that $\hat\Phi_i$ and $\hat\Pi_i$ are real, we have
by Hermicity that \be\label{hermit} \hat x^{\dag }_i({\bf k},t)=\hat
x_i(-{\bf k},t),\;\;\; \hat \pi^{\dag}_i({\bf k},t)=\hat\pi_i(-{\bf
k},t), \ee
and we also take $\omega_i({\bf k}) =\omega_i(-{\bf k}) $.\\
Next let
\begin{eqnarray}\label{fock}
\hat x_i ({\bf k},t)&=& \sqrt{\frac{\hbar}{2\omega_i({\bf k})}}\left(\hat a_i({\bf k},t)+\hat a^{\dag}_i({-\bf k},t)\right),\nonumber\\
\hat\pi_i ({\bf k},t)&=&i \sqrt{\frac{\hbar\omega_i({\bf
k})}{2}}\left(\hat a^{\dag}_i({\bf k},t)-\hat a_i({-\bf
k},t)\right).
\end{eqnarray}
It readily follows from (\ref{fock}) and (\ref{cancomm}) that
\be
[{\hat a}_i({\bf k},t), {\hat a}^{\dag}_j({\bf k'},t)]=\delta_{ij}\delta({\bf k}-{\bf k'}),\nonumber
\ee
\be
[{\hat a}_i({\bf k},t), {\hat a}_j({\bf k'},t)]=0,
\ee
\be
[\hat a^{\dag}_i({\bf k},t), \hat a^{\dag}_j({\bf k'},t)]=0.\nonumber
\ee
So $\hat a^{\dag}_i $ and $\hat a_i $, are the usual Fock creation and destruction operators. Note
however that while particle-antiparticle degeneracy at the dispersion relation level is
preserved for a given value of the label $i=1,2$, the energies of the particles-antiparticles  created (destroyed)
by $\hat a^{\dag}_i $ ($\hat a_i $) are different and are given by $\hbar\omega_i$.\\

\section{Discussion and Conclusions}

Spacetime noncommutativity in field theory is understood in some circles
as a merely convenient way to describe a special type of interaction.
Such a description consisting in mathematically deforming the product in the algebra of field
functions by means of the so-called Moyal star-product. However, as we tried to stress
throughout the paper, referring to the formalism under such premises as
spacetime noncommutativity is, at best, a misnomer since the arguments of the fields are parameters of the theory.
Speaking about noncommutativity in this context then has
little physical basis, beyond the rather loose analogy of the Moyal product with the Groenewold- Moyal product occurring
in the WWGM phase-space formulation of quantum mechanics.
 One of our contentions here has been, however, that there is
more physical substance to that designation if one recalls the operational nature
of observables in quantum mechanics from where
noncommutativity of the dynamical
variables of the system is readily understood then as the noncommutatitivity of their corresponding operators.
Furthermore, based on the concept that quantum mechanics can be viewed as a
minisuperspace sector of field theory, where only a few degrees of
freedom are unfrozen, we have used the quantum
mechanics of a harmonic oscillator over an extended Heisenberg
algebra, to construct a field theoretical model which inherits the
space-space noncommutativity of the quantum mechanical problem.\\
An interesting feature of our construction is that it shows that the
global symmetry of the original theory (\ref{qham}) is broken by the
noncommutativity. This in turn implies that if at the level of field
theory the index tagging the fields denotes a composite system of
scalar fields (and not the components of a vector field), then the
noncommutativity can be seen as giving rise to a field doublet (or
more generally an n-tuplet) of slightly different masses where
classical Lorentz symmetry for each member is broken, but each one
satisfies a deformed Klein-Gordon equation which is invariant under
a twisted Lorentz symmetry. On the other hand, if the labeling of
the fields is taken as corresponding to that of a vector field of
spacetime dimensions then, because of the mass differences, both
classical and twisted Lorentz invariance are broken by the noncommutativity.\\
This symmetry breaking and mass differences resulting from the  presence of
noncommutativity is in some way reminiscent of the spontaneous symmetry breaking
mechanism that occurs in the Standard Model, but without the appearance of a Goldstone
boson.

In addition, by thinking of noncommutativity of spacetime as the
quantum mechanical operator algebra expressing the loss of operational
meaning for localization at distances of orders smaller than the
Planck length, it then follows that minisuperspaces based on noncommutative spacetimes
have to be at least of two dimensions, and the fields constructed from
them must necessarily contain the presence of the parameter of
noncommutativity even in the absence of self-interacting potentials.

An alternate way to mathematically express the physical argument that measurements  below
distances of the order of the Planck length loose operational significance, can be accomplished,
both at the quantum mechanical and field theoretical level, by using parametrization invariance of
the action and following the canonical quantization approach of embedding a spatial manifold $\Sigma$
in the spacetime manifold. Such an approach, whereby the embedding variables acquire a dynamical
interpretation, which, in turn, gives physical sense to their noncommutativity
and is achieved by the inclusion of a general symplectic structure in the formalism, has been
analyzed extensively by the authors elsewhere \cite{ros2}. The deformed algebra of the
constraints resulting from the parametrization and general symplectic structure of the theory
is particularly convenient for analyzing the twisting of its symmetries and for indeed
thinking of a true  physical  spacetime  noncommutativity  as underlying the merely  axiomatic
mathematical  deformation of the algebra product  describing a certain type of
interactions in field theory.

Finally, we note that although our construction has been restricted
for simplicity to two spacial dimensions and to bosonic fields, it can
be generalized to allow for higher dimensional spaces in a conceptually
straightforward (albeit algebraically more complicated) way, and to the
case of fermionic fields by including Grassmanian variables in the
construction of the spectral oscillators.

\section*{Acknowledgements} The authors acknowledge partial support from
CONACyT project UA7899-F (M. R.),  DGAPA-UNAM grant IN104503-3
(J.D.V.) and SEP-CONACyT project 47211-F  (J.D.V).

\end{document}